# Comparing like with like: China ranks first in SCI-indexed research articles since 2018

Junwen Zhu

jwzhu@ed.ecnu.edu.cn

Faculty of Education, East China Normal University, Shanghai 200062, China

Weishu Liu    Corresponding author

wsliu08@163.com

School of Information Management and Engineering, Zhejiang University of Finance and Economics, Hangzhou 310018, Zhejiang, China

**Abstract:** China's rising in scientific research output is impressive. The academic community is curious about the time when the cross-over in the number of annual scientific publication production between China and the USA can happen. By using Web of Science Core Collection's Science Citation Index Expanded database, this study finds that China still ranks the second in the production of SCI-indexed publications in 2019 but may leapfrog the USA to be the first in 2020 or 2021, if all document types are considered. Comparatively, China has already overtaken the USA and been the largest SCI-indexed original research article producer since 2018. However, China still lags behind the USA regarding the number of review paper production. In general, quantitative advantage does not equal quality or impact advantage. We think that the USA will continue to be the global scientific leader for a long time.

## Introduction

China, the largest developing economy, is attracting increasing attention globally (Cao & Suttmeier 2017; Liu et al. 2015b). China's rapid rising in scientific research is also impressive (Liu et al. 2015a; Nature Editorial 2020; Tollefson 2018; Wang 2016; Zhou & Leydesdorff 2006). Based on the data from Web of Science, many previous studies have found that China has been the second largest producer of scientific publications for over ten years (Leydesdorff & Wagner 2009; Tang 2019; Zhou & Leydesdorff 2008).

Academia is curious about the time when China can overtake the USA as the largest producer of scientific publications (Leydesdorff 2012; Zhou 2013). However, the result is database dependent. For example, Nature has reported that China has been the largest producer of scientific publications by using Elsevier's Scopus database (Tollefson 2018), which is different from the finding based on Web of Science. Besides, different document types vary significantly regarding the academic value. However, previous studies tend to take into account all the document types or some important document types, such as articles and reviews. One study has shown that different countries may have quite different document-type country profiles (Zhang et al. 2011). Therefore, it is more suitable to compare like with like: compare apples to apples rather than to oranges.

## Data and methods

In this study, the Science Citation Index Expanded (hereinafter SCI for brevity) under the Web of Science Core Collection platform was chosen as the data source (Liu 2019; Zhu & Liu 2020). We restricted a 20-year timespan from 2000 to 2019 for this study. The whole counting method was used. For example, if a paper is co-authored by four authors from three different countries, all these three unique countries will get one full credit. We merge England, Scotland, Wales, and North Ireland into the United Kingdom. Since article and review are two most important document types in terms of the academic value, this study will treat these two document types separately in scenario 2 and 3. The data were accessed on April 25, 2020 via the library of Xi'an Jiao Tong University.

## Document-type country profiles

Table 1 shows the document type distribution of the whole SCI database, publications from the United States, and publications from China during the past two decades. The top 10 document types in SCI database are listed in descending order. Marked difference exists among the three groups. Article is the prominent document type with the relative share of 70.24% in the SCI database, followed by meeting abstract (15.76%), editorial material (4.70%), review (4.22%), and proceeding paper (4.19%)[1].

The document type distribution of publications contributed by the United States is a bit similar to the world total. During the past two decades, 64.45% of the SCI-indexed publications are articles, followed by meeting abstracts (21.69%), editorial materials (5.50%), reviews (5.13%), and proceeding papers (3.87%). That is to say, compared with the world average, United States' SCI-indexed publications demonstrate a bit lower share in original research articles but a bit higher share in meeting abstracts.

Table 1 Document-type country profiles

| Rank | Document type | World (#) | Relative share (%) | USA (#) | Relative share (%) | China (#) | Relative share (%) |
|---|---|---|---|---|---|---|---|
| 1 | Article | 21626789 | 70.24 | 5702391 | 64.45 | 3358394 | 91.15 |
| 2 | Meeting abstract | 4852261 | 15.76 | 1918890 | 21.69 | 142852 | 3.88 |
| 3 | Editorial material | 1445760 | 4.70 | 486550 | 5.50 | 32904 | 0.89 |
| 4 | Review | 1298427 | 4.22 | 453979 | 5.13 | 106281 | 2.89 |
| 5 | Proceedings paper | 1290117 | 4.19 | 342179 | 3.87 | 83227 | 2.26 |
| 6 | Letter | 788399 | 2.56 | 199697 | 2.26 | 27643 | 0.75 |
| 7 | News item | 390937 | 1.27 | 12556 | 0.14 | 2080 | 0.06 |
| 8 | Correction | 236574 | 0.77 | 29221 | 0.33 | 12961 | 0.35 |
| 9 | Biographical item | 71734 | 0.23 | 13722 | 0.16 | 355 | 0.01 |
| 10 | Book review | 64790 | 0.21 | 26245 | 0.30 | 169 | 0.01 |

Data source: Science Citation Index-Expanded (2000-2019)

---

[1] A proceeding paper covered in SCI database is also identified as an article when published in a journal. Please refer to: http://images.webofknowledge.com//WOKRS535R52/help/WOS/hs_document_type.html

Interestingly, the document type distribution of the SCI-indexed publications contributed by China is quite different from the previous two groups. Over 90% of China's SCI-indexed publications are original articles, which is much higher than the previous two groups. Contrarily, meeting abstracts only take a share of 3.88% among all China's publications, which is only one quarter of the world average. Lower presence rates also exist for many other document types. However, we should note the serious author address missing problems in Web of Science for some document types such as news item, correction, and biographical item (Liu, Hu, & Tang 2018).

**Scenario 1: all document types are taken into account**

Figure 1 shows the annual production volume of SCI publications of the United States and China during the past two decades when all document types are taken into account. According to Figure 1, the United States is the largest SCI publication contributor consistently during all the past 20 years. However, its annual production volume grew slowly from 322909 in 2000 to 574706 in 2019, with an increase of 78%.

However, China's annual production volume of SCI publications grew rapidly from 31114 in 2000 to 513435 in 2019 (an amazing increase of 1550%). Besides, the growth rate even seemed to increase in the past two years. As for the ranking, China has overtaken Germany and the United Kingdom as the second largest SCI publication contributor since 2008. We predict that China will overtake the United States as the largest SCI publication producer in 2020 or 2021, if all document types are considered.

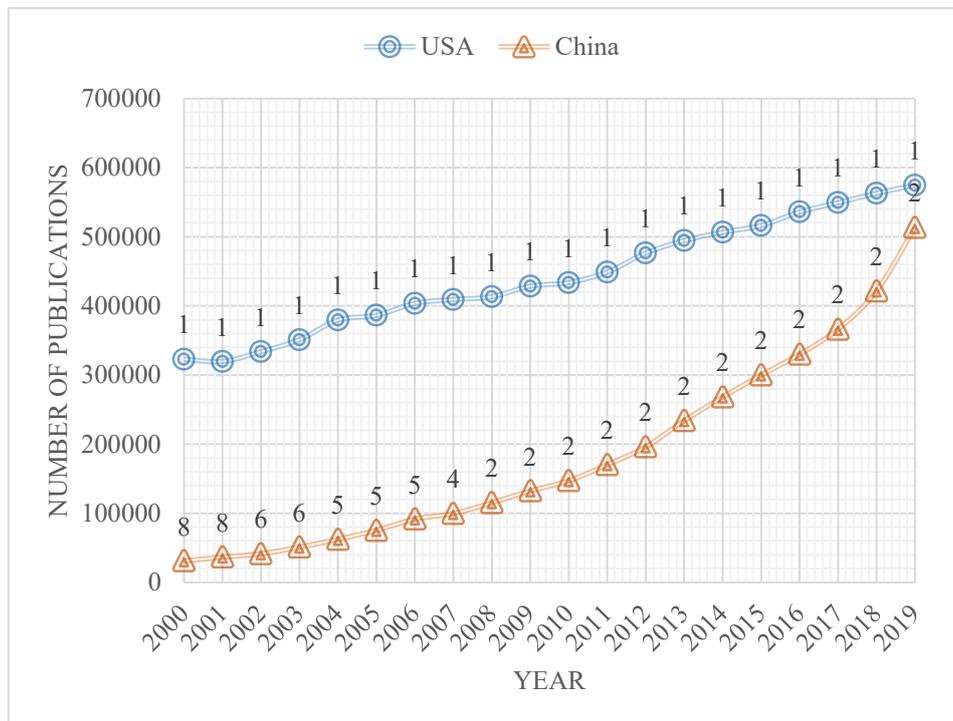

Figure 1 Numbers of SCI-indexed publications: the United States vs China

Note: all document types are considered.

Figure 2 also shows the relative shares of the world total SCI publications contributed by the United

States and China. Even though the absolute number of SCI publications produced by the United States kept increasing, its relative share remained declining from 33% in 2000 to 25% in 2019. Contrarily, China's relative share also rose rapidly from 3% in 2000 to 23% in 2019.

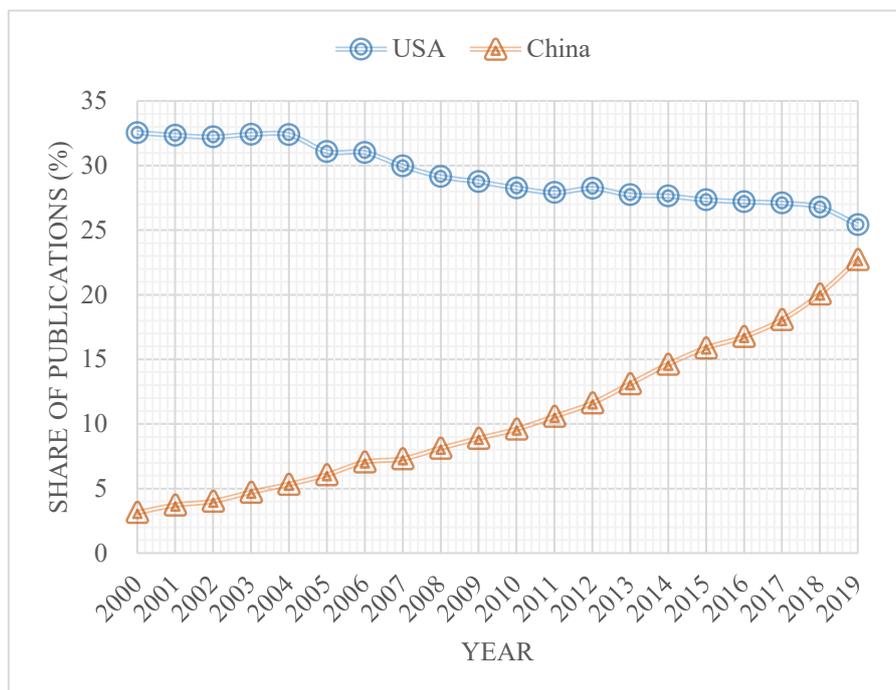

Figure 2 Relative shares of SCI-indexed publications: the United States vs China

Note: all document types are considered.

## Scenario 2: only articles are taken into account

Since the document type distribution for the SCI publications of the United States and China vary considerably, it is more suitable to compare apples to apples rather than to oranges. In this section, we only consider original research articles identified by Web of Science for comparison.

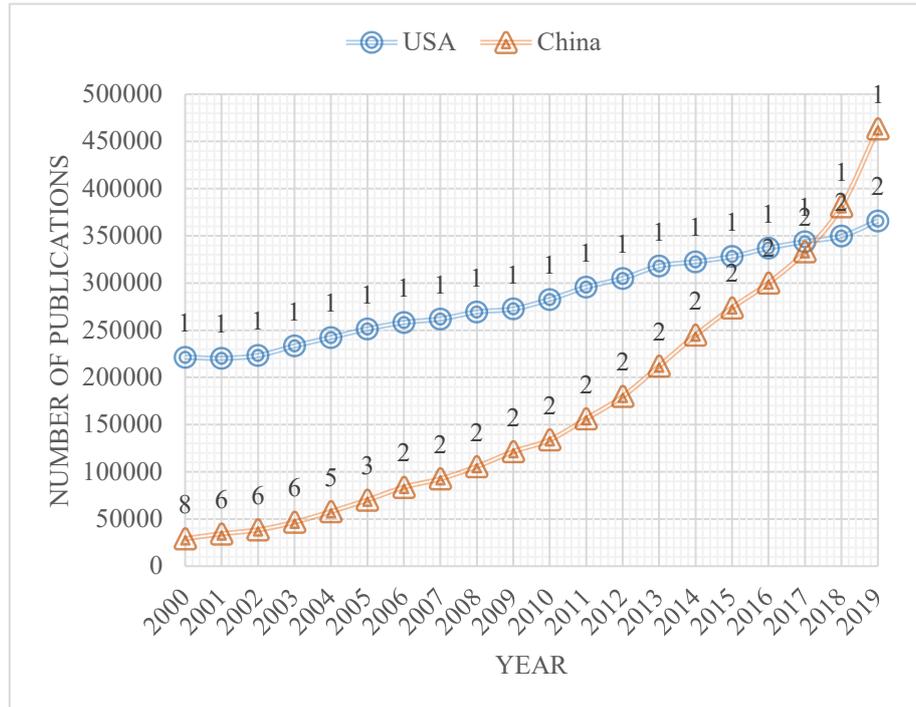

Figure 3 Numbers of SCI-indexed articles: the United States vs China
Note: only original research articles are considered.

Figure 3 demonstrates the annual production of SCI-indexed original research articles by the United States and China. Similarly, the number of SCI-indexed articles of the United States rose slowly from 221418 in 2000 to 365837 in 2019, with an increase of 65%. However, the corresponding number contributed by China rose rapidly from 29141 in 2000 to 463326 in 2019 with an increase of 1490%. Besides, China has been the second largest contributor to SCI-indexed articles since 2006, which is two years earlier than that in the scenario 1. Surprisingly, China has overtaken the United States as the largest contributor of original research articles in 2018. What's more, China's production of SCI-indexed research articles in 2019 is 27% higher than that of the United States.

Figure 4 demonstrates the gradual decrease of the relative contribution by the United States and the rapid increase of the relative share contributed by China. More specifically, the relative share of the United States decreased from 31% in 2000 to 22% in 2019. Comparatively, China's share rose from 4% in 2000 to 28% in 2019.

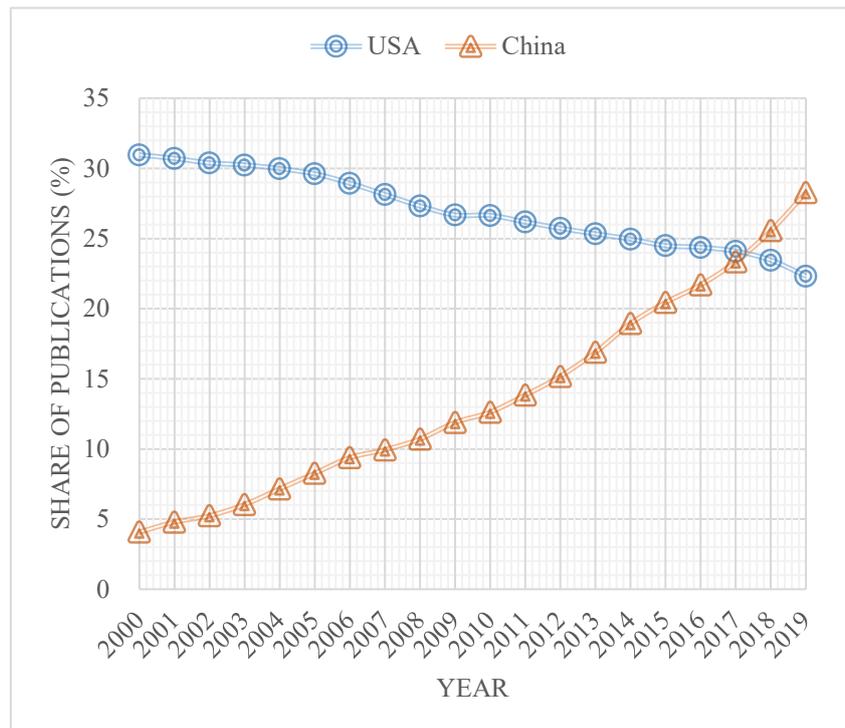

Figure 4 Relative shares of SCI-indexed articles: the United States vs China

Note: only original research articles are considered.

**Scenario 3: only reviews are taken into account**

Article and review are two substantial citable items in Web of Science. Apart from scenarios 1 and 2, we further only take reviews into account in scenario 3.

Figure 5 shows the annual production of review papers by the United States and China during the past two decades. The United States has remained a moderate growth trend and been the largest producer of review papers consistently over the last 20 years. The SCI-indexed reviews for the United States rose from 12957 in 2000 to 37226 in 2019, with an increase of 187% which is much higher than its corresponding increases in previous two scenarios. As for the review papers, China only published 328 papers in 2000 and ranked 18[th] among all the countries. The absolute number of China's review papers also grew slowly during the first decade but accelerated during the second decade. Besides, China has been the second largest contributor to SCI-indexed reviews since 2016, which is eight and ten years later than that in the scenario 1 and 2, respectively. Maybe it will still take several years for China to overtake the United States as the largest review paper producer.

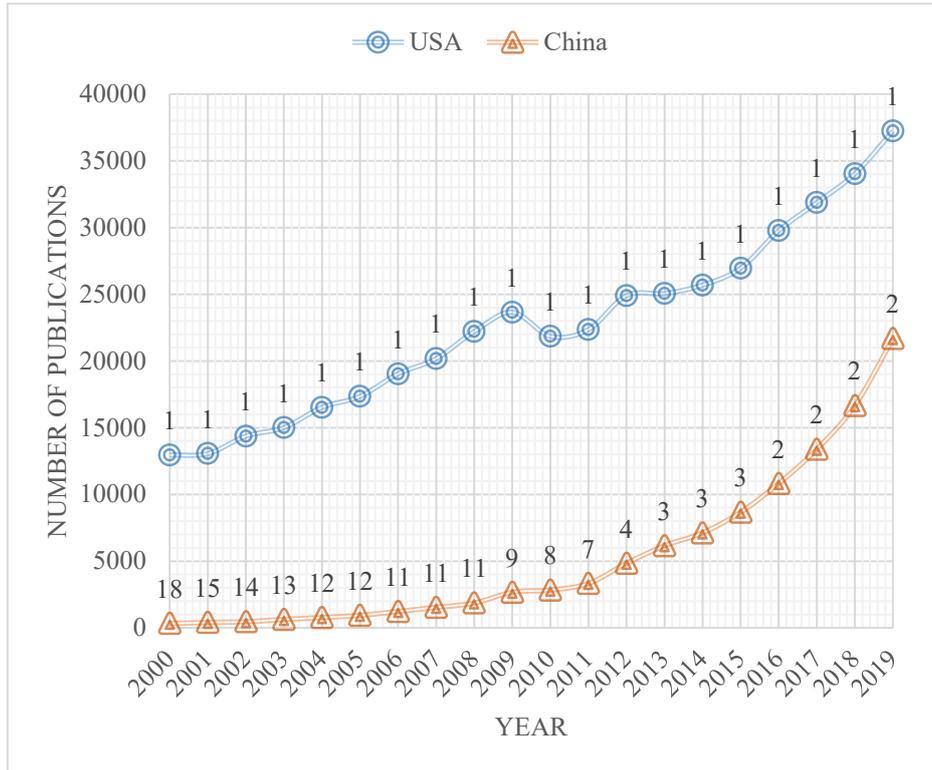

Figure 5 Numbers of SCI-indexed reviews: the United States vs China

Figure 6 also demonstrates the relative shares of the contribution of the United States and China to world total SCI-index reviews during the past two decades. Similarly, the declining trends of the relative shares of the United States and the increasing trends of the relative shares of China co-exist. The relative share of the United States decreased from 43% in 2000 to 29% in 2019, however, the relative share of China increased from 1% in 2000 to 17% in 2019.

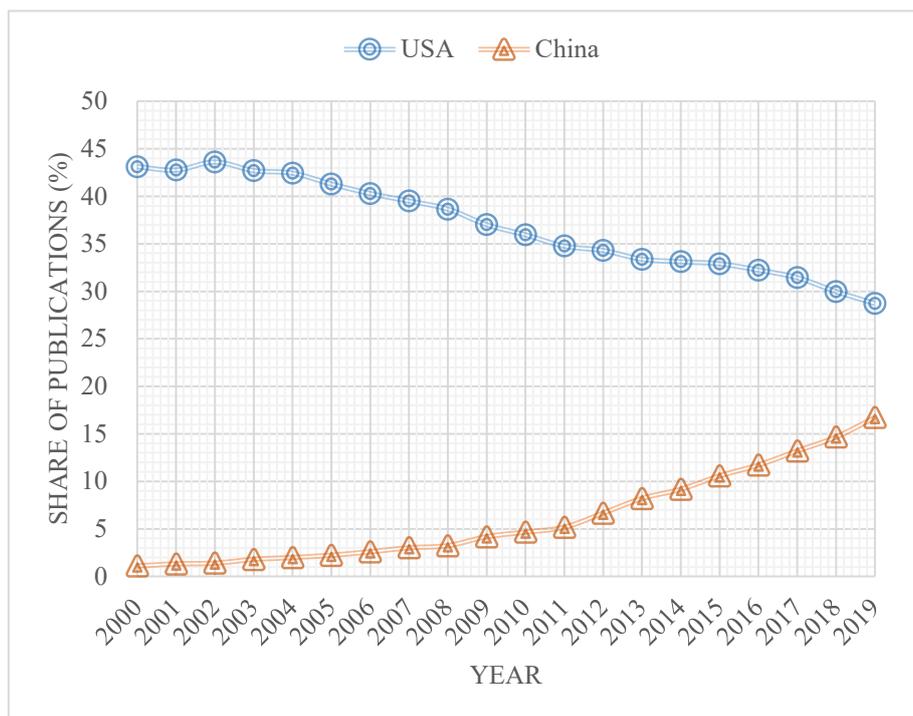

Figure 6 Relative shares of SCI-indexed reviews: the United States vs China

## Discussion

This study re-check the fact that the United States and China have quite different document type country-profiles which has been documented by Zhang et al. (2011). Different document types vary significantly, therefore it is not suitable to treat different document types equally. In this study, we divided the analysis into three different scenarios. If all the document types are considered, China is still the second largest producer of SCI-index publications but may overtake the United States in 2020 or 2021. However, China has overtaken the United States as the largest producer of SCI-indexed original research articles in 2018. Similar result can also be concluded by combining articles and reviews together (Liu 2020). Differently, despite China's status as the second largest producer of reviews since 2016, there is still a large gap in review paper production between the United States and China.

China's rapid rising in scientific research is widely documented (Basu et al. 2018; Quan et al. 2019). And China has even overtaken the United States as the largest producer of SCI-indexed original research articles in 2018. However, the quality/impact of China's scientific research outputs still lags behind, and a certain extent of paper bubbles also exists in China (Leydesdorff et al. 2014; Liu 2020). China needs to shift its research evaluation standards to emphasize more on the improvement of research quality and halt some problematic publishing practices (Liu 2020; Tang 2019; Tang et al. 2020).

Due to the COVID-19 pandemic, the growth of scientific output of the United States and China will possibly slow down and even turn negative. The latest reform in China may also decelerate the rapid growth of China's SCI publications (Liu 2020). However, the number of junior researchers who can publish in English is also increasing. We still predict that China will overtake the United States as the largest contributor to SCI-indexed publications in the following one or two years (all document types are considered). However, the cross-over between China and the United States cannot be misinterpreted or overinterpreted. The United States will continue to be the global scientific leader for a long time (Tollefson 2018).

**Compliance with ethical standards**

Conflict of interest The authors declare that there is no conflict of interest.